\documentclass{l4dc2024}

\usepackage{amssymb}
\usepackage{amsmath}
\usepackage{latexsym}
\usepackage{graphicx}
\usepackage{color}
\usepackage{url}
\usepackage{Shorthands}
\usepackage{caption}

\usepackage{listings}
\usepackage{courier}
\usepackage{comment}
\setcitestyle{square}
\SetKwInput{KwInput}{Input}                
\SetKwInput{KwOutput}{Output}              

\usepackage{algorithm,algcompatible}
\algnewcommand\INPUT{\item[\textbf{Input:}]}%
\algnewcommand\OUTPUT{\item[\textbf{Output:}]}%

\usepackage{hyperref}

\renewcommand{\hat}{\widehat}
\newcommand{\p}{{\mathbf p}}

\usepackage{mathtools}

\newcommand{\qed}{\hfill\blacksquare}
\makeatletter
\renewcommand*{\@opargbegintheorem}[3]{\trivlist
  \item[\hskip \labelsep{\bfseries #1\ #2}] \textbf{(#3)}\ \itshape}
\makeatother

\newtheorem{mytheorem}{Theorem}
\newtheorem{mylemma}{Lemma}
\newtheorem{myremark}{Remark}
\newtheorem{myassumption}{Assumption}
\newtheorem{myproblem}{Problem}

\newcommand{\Yhat}{\ensuremath{\hat{Y}}}

\title[Recursively Feasible MPC in Dynamic Settings with Conformal Prediction Guarantees]{Recursively Feasible Shrinking-Horizon MPC in Dynamic Environments with Conformal Prediction Guarantees}
\usepackage{times}

\author{%
 \Name{Charis Stamouli}$^1$ \Email{stamouli@seas.upenn.edu}\\
 \Name{Lars Lindemann}$^2$ \Email{llindema@usc.edu}\\
 \Name{George J. Pappas}$^1$ \Email{pappasg@seas.upenn.edu}\\
 \addr $^1$University of Pennsylvania, Philadelphia, PA\\
 \addr $^2$University of Southern California, Los Angeles, CA 
}

\begin{document}

\maketitle

\captionsetup[figure]{labelfont={bf},labelformat={default},labelsep=colon,name={Fig.}}

\begin{abstract}
In this paper, we focus on the problem of shrinking-horizon Model Predictive Control (MPC) in uncertain dynamic environments. We consider controlling a deterministic autonomous system that interacts with uncontrollable stochastic agents during its mission. Employing tools from conformal prediction, existing works derive high-confidence prediction regions for the unknown agent trajectories, and integrate these regions in the design of suitable safety constraints for MPC. Despite guaranteeing probabilistic safety of the closed-loop trajectories, these constraints do not ensure 
feasibility of the respective MPC schemes for the entire duration of the mission. We propose a shrinking-horizon MPC that guarantees recursive feasibility via a gradual relaxation of the safety constraints as new prediction regions become available online. This relaxation enforces the safety constraints to hold over the least restrictive prediction region from the set of all available prediction regions. 
In a comparative case study with the state of the art, we empirically show that our approach results in tighter prediction regions and verify recursive feasibility of our MPC scheme.
\end{abstract}


\section{Introduction}
In safety-critical situations, predicting the behavior of uncontrollable dynamic agents is essential for systems that operate among them. For example, a mobile robot that navigates among pedestrians is responsible for the safety of itself and any humans it encounters on its path. To facilitate planning in such environments, Model Predictive Control (MPC) schemes that incorporate predictions of the future agent states have been developed, see, e.g., \citep{Trautman2010,Kuderer2012,Wang2022,Yoon2021,Nair2022}. However, uncertainty in the predictions can jeopardize safety, potentially allowing the system to approach the true location of the agents.

The authors in \citep{Du2011,Du2011old} quantify prediction uncertainty in the case of spherical or Gaussian-distributed agents and derive a probabilistic collision-checking constraint for MPC. In \citep{Fisac2018,Fridovich2020}, Bayesian inference is used to estimate uncertainty in the prediction of human motion. The scheme in \citep{Zhu2022} provides a sample-based estimate of the uncertainty for Gaussian-process predictors. In \citep{Wei2022}, a bootstrapped predictor is employed and distribution-free constraints are derived to ensure probabilistic safety. In all the aforementioned works, uncertainty quantification is restricted to particular predictors (e.g., Gaussian processes) or agent models/distributions (e.g., Gaussian agents). These limitations prevent the corresponding MPC schemes from providing safety guarantees for state-of-the-art predictors (e.g., neural networks) and arbitrary stochastic agents (e.g., non-Gaussian agents). 

In statistics, a more general technique for uncertainty quantification is provided by conformal prediction \citep{Vovk2005,Shafer2008}. The key idea is to use a calibration dataset to design high-confidence prediction regions for the test data. This method is compatible with any prediction algorithm, without imposing any assumptions on the data distribution. Conformal prediction has been recently employed in various safety-critical applications of autonomous systems. Application examples include detection of unsafe situations \citep{Luo2022}, generation of probabilistically safe reachable sets \citep{Muthali2023}, safety verification \citep{Bortolussi2019,Fan2020,Lindemann2023b}, and safe open-loop control \citep{Tonkens2023}.

In the context of MPC, conformal prediction for dynamic agent trajectories has been explored to certify safety during control implementation. When the test trajectories are included in the training data, the conformal prediction regions from \citep{Chen2021} guarantee probabilistic safety. The MPC scheme in \citep{Dixit2023} adaptively quantifies prediction uncertainty, providing an average probabilistic safety guarantee. Conceptually closest to our work are the  shrinking-horizon MPC schemes presented in \citep{Lindemann2023a,Yu2023}, where safety is ensured jointly at all time steps with high probability. In \citep{Lindemann2023a}, the authors achieve this by union bounding over the one-step ahead prediction regions corresponding to all time steps. In contrast, the authors in \citep{Yu2023} simultaneously design high-confidence regions for one-step ahead predictions without employing union bounding. Despite their strong safety guarantees, neither of the respective MPC schemes is ensured to be feasible for the entire duration of system operation. 

\begin{figure}
   \centering
   \includegraphics[width=0.5\textwidth]{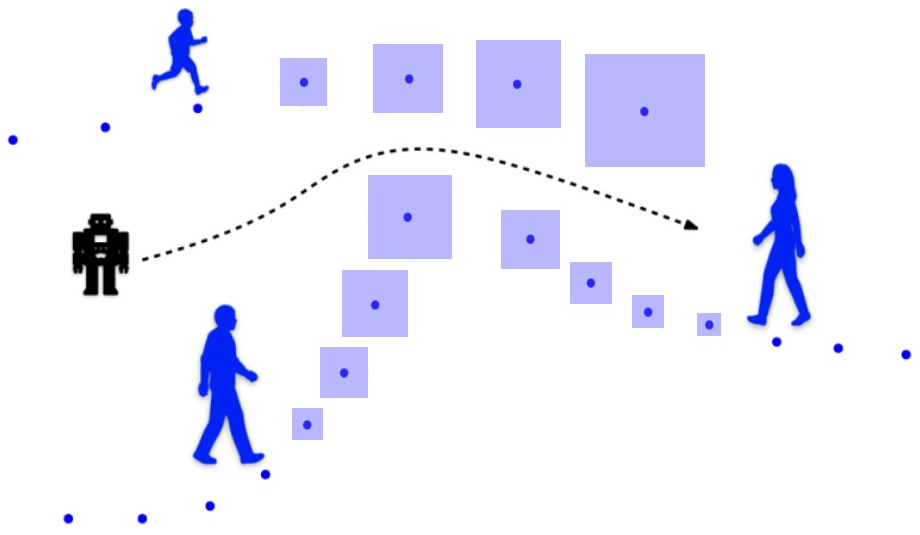}
   \caption{{\small We predict trajectories of dynamic agents using arbitrary predictors (e.g., neural networks) and employ conformal prediction to bound the uncertainty in high-confidence prediction regions (blue squares).}}
   \label{fig_intro}
\end{figure}

In this paper, we focus on jointly providing recursive feasibility and probabilistic safety guarantees for shrinking-horizon MPC in dynamic environments. In particular, we consider a deterministic autonomous system interacting with uncontrollable dynamic agents, whose trajectories follow an unknown distribution. Given a learning-based trajectory predictor, we obtain online-updated estimates of the agent trajectories. Employing these estimates along with tools from conformal prediction, we derive updated prediction regions for the future agent states at each time step (see Figure~\ref{fig_intro}). Unlike \citep{Lindemann2023a}, where union bounding is employed, we simultaneously construct high-confidence prediction regions over multiple time steps, inspired by the approach in \citep{Yu2023}. In contrast to \citep{Lindemann2023a,Yu2023}, where high-confidence regions for one-step ahead predictions are derived, we obtain high-confidence prediction regions for all prediction time steps. We leverage these regions to design appropriate safety constraints for all future system states at each time step. Our main contributions are the following:
\vspace*{-0.2cm}
\begin{enumerate}[label=\roman*)]
\item Employing our safety constraints, we develop a recursively feasible and probabilistically safe shrinking-horizon MPC scheme. Recursive feasibility is ensured by the gradual relaxation of our constraints as updated prediction regions are constructed online. Probabilistic safety is guaranteed by the fact that our constraints rely on high-confidence prediction regions. To the best of our knowledge, our MPC scheme is the first to ensure recursive feasibility, while maintaining probabilistic safety, for arbitrary trajectory predictors and stochastic agents.
\vspace*{-0.35cm}
\item We empirically show that our conformal prediction method results in  tighter prediction regions than the one in \citep{Lindemann2023a}. We also verify recursive feasibility of the proposed MPC scheme in a comparative simulation.
\end{enumerate}
\vspace*{-0.3cm}
All proofs are included in Appendix~\ref{appendix_A}.

\section{Problem Formulation}\label{Problem_Formulation}
Consider a discrete-time nonlinear dynamical system of the form:
\begin{equation}\label{system}
    x_{t+1} = f(x_t,u_t),
\end{equation}
where $x_t\in\setR^n$ denotes the state, $u_t\in\setR^m$ the control input, and $f:\setR^n\times\setR^m\to\setR^n$ the dynamics function. Suppose that the system is subject to state and input constraints of the form:
\begin{align}
    \label{state_constraints}
    &x_t\in\calX_t,\; t=0,\ldots,T,\\
    \label{input_constraints}
    &u_t\in\calU_t,\; t=0,\ldots,T-1,
\end{align}
respectively, where $T\geq1$ is the total mission time.\footnote{We assume that the initial state satisfies the constraint \eqref{state_constraints} (i.e., $x_0\in\calX_0$).} Moreover, consider a control performance objective determined by the following cost function:
\begin{equation}\label{cost_function}
    J_T(x_0;u_0,\ldots,u_{T-1}) = J_f(x_T)+\sum_{t=0}^{T-1}J(x_t,u_t),
\end{equation}
where $J:\setR^n\times\setR^m\to\setR_+$ evaluates the stage cost at each time step $t=0,\ldots,T-1$, and $J_f:\setR^n\to\setR_+$ the terminal cost at time $T$.

The system operates in an environment with $N$ dynamic stochastic agents whose trajectories are a priori unknown. Let $Y_t:=\begin{bmatrix}Y_{t,1}^{\intercal},\ldots,Y_{t,N}^{\intercal}\end{bmatrix}^\intercal$ represent the joint agent state at time $t$, where $Y_{t,j}\in\setR^{p}$ denotes the state of agent $j$ at time $t$. The system can sequentially collect observations $Y_0,\ldots,Y_t$ of the joint agent state, while being controlled up to time $t$. The joint agent trajectory $Y_{0:T}:=\begin{bmatrix}Y_0^{\intercal},\ldots,Y_T^{\intercal}\end{bmatrix}^{\intercal}$ is assumed to be sampled from an unknown probability distribution $\calD$. 

\begin{myassumption}\label{assumption_distribution_shifts}
The state $x_t$ of system \eqref{system} does not change the distribution $\calD$ of the joint agent trajectory $Y_{0:T}$, for any $t=0,\ldots,T-1$.
\end{myassumption}

Assumption~\ref{assumption_distribution_shifts} approximately holds in many applications and is standard in prior work \citep{Lindemann2023a,Yu2023}. 
We reserve a comprehensive study of distribution shifts for future work. In addition to the constraints \eqref{state_constraints} and \eqref{input_constraints}, we consider the dynamic agent constraints:
\begin{equation}\label{agent_constraints}
    c(x_t,Y_t)\geq0,\;t=0,\ldots,T,
\end{equation}
where $c:\setR^n\times\setR^{Np}\to\setR$ is $L$-Lipschitz in its second argument with respect to some norm $\norm{\cdot}$.\footnote{\label{footnote2}We assume that the initial state satisfies the constraint \eqref{agent_constraints} (i.e., $c(x_0,Y_0)\geq0$).} While safety is the main focus of our work, the function $c(\cdot,\cdot)$ can encode diverse objectives, from collision avoidance to dynamic agent tracking (see Section~\ref{Case_Study} for a particular example of collision avoidance). Given the fact that the trajectory $Y_{0:T}$ is a priori unknown, the constraints imposed by \eqref{agent_constraints} on the system state $x_t$ are also a priori unknown. However, we assume availability of an offline dataset $D:=\{Y_{0:T}^{(1)},\ldots,$ $Y_{0:T}^{(K)}\}$, where the joint agent trajectories $Y_{0:T}^{(i)}:=\begin{bmatrix}Y_0^{{(i)}^\intercal},\ldots,Y_T^{{(i)}^\intercal}\end{bmatrix}^{\intercal}$ are independently drawn from $\calD$. Employing the dataset along with online observations of $Y_{0:T}$, our goal is to develop a recursively feasible MPC with probabilistic safety guarantees for system \eqref{system}.

\begin{myproblem}[Recursively Feasible and Probabilistically Safe Shrinking-Horizon MPC in Dynamic Environments]\label{problem}Given  a failure probability $\delta\in(0,1)$, design a recursively feasible shrinking-horizon model predictive controller for system \eqref{system}, that minimizes the cost function \eqref{cost_function}, while guaranteeing the satisfaction of \eqref{state_constraints}, \eqref{input_constraints}, and:
\begin{equation}\label{prob_agent_constraints}
    \Prob\left(\bigcap_{t=0}^T\big\{c(x_t,Y_t)\geq0\big\}\right)\geq1-\delta.
\end{equation}
\end{myproblem}

In the next section, we provide some background on trajectory predictors and employ conformal prediction to derive high-confidence prediction regions for the future agent states, for every $t=0,\ldots,T-1$. We will leverage these regions to develop a recursively feasible shrinking-horizon MPC scheme that ensures satisfaction of the probabilistic safety guarantee \eqref{prob_agent_constraints} (see Section~\ref{MPC_in_Dynamic_Environments_via_Conformal_Prediction}).

\section{Online-updated Conformal Prediction Regions for Dynamic Agent Trajectories}\label{Conformal_Prediction_Regions_for_Dynamic_Agent_Trajectories}
The main challenge in addressing Problem~\ref{problem} lies in the absence of prior knowledge of the joint agent trajectory $Y_{0:T}$. The trajectory $Y_{0:T}$ is a key component of the dynamic agent constraints \eqref{agent_constraints}, which are imposed on the state of system \eqref{system}, and are critical to the safety of a model predictive controller. To address this issue, we first split the dataset $D$ into  a training set $D_{\train}$ and a calibration set $D_{\calib}$. Employing the dataset $D_{\train}$, we can develop an algorithm that, at each time $t$, predicts the future agent states $Y_{t+1},\ldots,Y_T$, as described in Subsection~\ref{Trajectory_Predictors}. Exploiting the dataset $D_{\calib}$, we can bound the uncertainty of the resulting estimates in high-confidence conformal prediction regions, as detailed in Subsection~\ref{Conformal_Prediction_Regions_based_on_Trajectory Predictors}. We note that our results in Subsection~\ref{Conformal_Prediction_Regions_based_on_Trajectory Predictors} are inspired by the approach in \citep{Yu2023} (see Remark~\ref{remark1} for details).

\subsection{Trajectory Predictors}\label{Trajectory_Predictors}
Recall that system \eqref{system} can sequentially collect observations $Y_0,\ldots,Y_t$ of the joint agent state, while being controlled up to time $t$. By feeding $Y_t$ into a trajectory prediction algorithm, we can generate estimates $\Yhat_{t+1|t},\ldots,\Yhat_{T|t}$ of the future agent states $Y_{t+1},\ldots,Y_T$, respectively. 

As outlined in \citep{Lindemann2023a}, a trajectory predictor can be formed by learning a predictive model $g:\setR^{Np}\to\setR^{Np}$ from the dataset  $D_{\train}$. For a given observation $Y_t$, the model $g(\cdot)$ should produce an estimate $\Yhat_{t+1|t}$ of the next state $Y_{t+1}$. We can recursively generate the remaining estimates $\Yhat_{t+2|t},\ldots,\Yhat_{T|t}$, by sequentially inputting $\Yhat_{t+1|t},\ldots,\Yhat_{T-1|t}$ to the learned model $g(\cdot)$. This method is compatible with any learning algorithm and model architecture. For example, the function $g(\cdot)$ could be modeled as a Recurrent Neural Network (RNN), an architecture that has been widely adopted in trajectory prediction \citep{Salehinejad2017}.

\subsection{Online-updated Conformal Prediction Regions based on Trajectory Predictors}\label{Conformal_Prediction_Regions_based_on_Trajectory Predictors}
In the previous subsection, we explained that at each time $t$, we can obtain predictions $\Yhat_{t+1|t},\ldots,$ $\Yhat_{T|t}$ of the future agent states $Y_{t+1},\ldots,Y_T$, respectively, by using trajectory predictors. To bound the uncertainty of these predictions, we will design valid prediction regions of the form:
\begin{equation}\label{prediction_regions}
    \norm[\big]{Y_{\tau}-\Yhat_{\tau|t}}\leq C_{\tau|t}.
\end{equation}
By valid prediction regions, we mean that for any given $\delta\in(0,1)$, the following guarantee:
\begin{equation}\label{conformal_prediction_guarantee}
    \Prob\left(\bigcap_{t=0}^{T-1}\bigcap_{\tau=t+1}^T\left\{\norm[\big]{Y_{\tau}-\Yhat_{\tau|t}}\leq C_{\tau|t}\right\}\right)\geq1-\delta
\end{equation}
should be satisfied by \eqref{prediction_regions}.\footnote{Note that the values $C_{\tau|t}$ may depend on $Y_{0:T}^{(1)},\ldots,Y_{0:T}^{(K)}$. Consequently, the probability measure $\Prob(\cdot)$ is defined over the product measure of 
$Y_{0:T}$ and $Y_{0:T}^{(1)},\ldots,Y_{0:T}^{(K)}$.} 

The challenges in deriving valid prediction regions are twofold. First, the distribution $\calD$ over trajectories is unknown and may deviate from standard assumptions (e.g., Gaussianity). Second, the trajectory predictor can be highly complex (consider, e.g., a neural network predictor). To address these issues, we employ a tractable variant of conformal prediction, referred to as split conformal prediction \citep{Papadopoulos2008}. This method is compatible with any prediction algorithm, without imposing any assumptions on the data distribution. Let $Y_t^{(i)}$ denote the state at time $t$ in the trajectory $Y_{0:T}^{(i)}\in D$ and let $\Yhat_{\tau|t}^{(i)}$ be the prediction of $Y_{\tau}^{(i)}$ at time $t$. Moreover, let $\calI_{\train}=\{i:Y_{0:T}^{(i)}\in D_{\train}\}$ and $\calI_{\calib}=\{i:Y_{0:T}^{(i)}\in D_{\calib}\}$. The main idea is to compute the values $C_{\tau|t}$ based on suitably defined random variables $R^{(i)}$, $i\in\calI_{\calib}$, which are called conformity scores. Conformity scores are typically determined by the prediction error on the calibration data. In standard supervised learning, they are usually given by $R^{(i)}=\norm{Z^{(i)}-\hat{Z}^{(i)}}$, where $\hat{Z}^{(i)}$ is the prediction of a calibration point $Z^{(i)}$. Herein, they are defined as:
\begin{equation}\label{conformity_scores}
    R^{(i)} = \max_{\substack{t=0,...,T-1\\\tau=t+1,...,T}}\left\{\frac{\norm[\big]{Y_{\tau}^{(i)}-\Yhat_{\tau|t}^{(i)}}}{\sigma_{\tau|t}}\right\},
\end{equation}
where:
\begin{equation}\label{normalization_factors}
    \sigma_{\tau|t}=\max_{j\in\calI_{\train}} \norm[\big]{Y_{\tau}^{(j)}-\Yhat_{\tau|t}^{(j)}},\;\forall t,\tau,
\end{equation}
for all $i\in\calI_{\calib}$. We can view the score \eqref{conformity_scores} as a normalized prediction error across all real time steps $t$ and prediction time steps $\tau$. Normalization via \eqref{normalization_factors} should approximately ensure that no single prediction error dominates the others in terms of scale.\footnote{For an alternative normalization, we could split the calibration data in two sets and compute the values $\sigma_{\tau|t}$ based on one of these sets.} This is crucial, because for any fixed $t$, the errors $\norm{Y_{\tau}^{(i)}-\Yhat_{\tau|t}^{(i)}}$ are expected to increase with the prediction time $\tau$. In the following lemma, we exploit the scores \eqref{conformity_scores} to design valid prediction regions of the form \eqref{prediction_regions}.

\begin{mylemma}[Online-updated Conformal Prediction Regions for Trajectories]\label{lemma1}
Fix a failure probability $\delta\in(0,1)$. Let $\Yhat_{\tau|t}$ be the prediction of the joint agent state $Y_{\tau}$ at time $t$. Let the conformity scores $R^{(i)}$ be as in \eqref{conformity_scores}, with normalizing factors $\sigma_{\tau|t}$ as in \eqref{normalization_factors}. Then,\, if\, $R$\, is\, the $\ceil{(|D_{\calib}|+1)(1-\delta)}$-th smallest value of the set $\{R^{(i)}: i\in\calI_{\calib}\}\cup\{\infty\}$, the guarantee \eqref{conformal_prediction_guarantee} holds with $C_{\tau|t}=R\sigma_{\tau|t}$, $\forall t,\tau$.\footnote{The notation $\ceil{\cdot}$ represents the ceiling function.}
\end{mylemma}

Notice in Lemma~\ref{lemma1} that the normalizing factors $\sigma_{\tau|t}$ are leveraged to suitably scale the score $R$ into the different values $C_{\tau|t}$. For the values $C_{\tau|t}$ to be finite, we need to have $\lceil(|D_{\calib}|+1)(1-\delta)\rceil\leq|D_{\calib}|$.

\begin{myremark}\label{remark1}
Conformal prediction regions of the form \eqref{prediction_regions} were previously derived in \citep{Lindemann2023a,Yu2023}. Therein, probabilistic guarantees were given only for one-step ahead predictions (i.e., for $\tau=t+1$, at each time step $t$). In Lemma~\ref{lemma1}, we extend the method of \citep{Yu2023} to provide valid prediction regions for all real time steps $t$ and prediction time steps $\tau$ (see \eqref{conformal_prediction_guarantee}). This is critical to simultaneously guaranteeing recursive feasibility and probabilistic safety of our MPC scheme (see Section~\ref{MPC_in_Dynamic_Environments_via_Conformal_Prediction}). Beyond that, empirical evidence suggests that our approach results in tighter prediction regions than the one in \citep{Lindemann2023a} (see Section~\ref{Case_Study}).
\end{myremark}

\section{Recursively Feasible Shrinking-Horizon MPC in Dynamic Environments with Conformal Prediction Guarantees}\label{MPC_in_Dynamic_Environments_via_Conformal_Prediction}
In this section, we present an MPC scheme that addresses Problem~\ref{problem}, exploiting the conformal prediction method described in Section~\ref{Conformal_Prediction_Regions_for_Dynamic_Agent_Trajectories}. Specifically, we focus on the case of shrinking-horizon MPC, where an optimal control problem of horizon extending to the end of the mission is solved at each time step. To ensure the probabilistic safety guarantee \eqref{prob_agent_constraints}, we derive reformulated dynamic agent constraints by leveraging the conformal prediction regions defined in Lemma~\ref{lemma1}. Recursive feasibility is guaranteed by the gradual relaxation of these constraints as updated prediction regions are constructed online. To achieve this relaxation, we progressively allow for states, that until the previous time step, were falsely predicted to be unsafe, as suggested by the latest prediction regions.

\begin{figure}[tbh]
   \centering
   \includegraphics[width=0.7\linewidth]{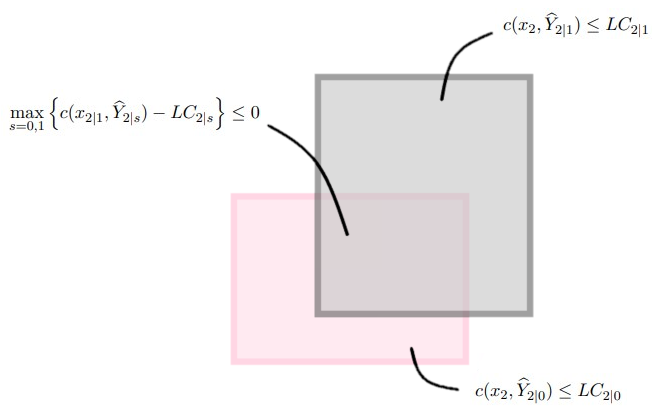}
   \caption{{\small Illustration of the constraint \eqref{relaxed_CP_agent_constraints} for collision avoidance. The predicted unsafe areas for $x_2$ at times $0$ (pink set) and $1$ (gray set) are guaranteed to be valid. Consequently, their intersection can be used to define the predicted unsafe area for $x_{2|1}$. This leads to an expanded area of free space for system \eqref{system} at time $1$.}}
   
   \label{fig_CP_shrinking}
\end{figure}

Before introducing our MPC scheme, we first derive the dynamic agent constraints that will be integrated in its design.
Let $t\in\{0,\ldots,T-1\}$ be the current time step, $\tau\in\{t+1,\ldots,T\}$ denote a future time step, and $s\in\{0,\ldots,t\}$ represent either a past or the current time step. Employing the conformal prediction regions of Lemma~\ref{lemma1}, we can show that for any given $\delta\in(0,1)$, with probability at least $1-\delta$, the following conditions jointly hold:
\begin{equation}\label{implication}
   c(x_{\tau},\Yhat_{\tau|t})\geq LC_{\tau|t} \implies c(x_{\tau},Y_{\tau})\geq0,\; \forall t,\tau.
\end{equation}  
The proof is omitted, as it is similar to the one of \citep[Theorem 3]{Lindemann2023a}. Owing to \eqref{implication}, we refer to the constraints: 
\begin{equation*}\label{multiple}
    c(x_{\tau},\Yhat_{\tau|t})\geq LC_{\tau|t},\;t=0,\ldots,\tau-1,
\end{equation*}
as \textit{valid predicted constraints} for $x_{\tau}$. We observe that at each time $t$, we have multiple valid predicted constraints for each future state $x_{\tau}$, which are given by:
\begin{equation}\label{multiple_prediction_regions}
    c(x_{\tau},\Yhat_{\tau|s})\geq LC_{\tau|s},\;s=0,\ldots,t.
\end{equation}
These constraints correspond to the prediction regions designed for $Y_{\tau}$ at time steps $s=0,\ldots,t$. Let $x_{\tau|t}$ be the prediction of $x_{\tau}$ made by a model predictive controller at time $t$. Since all constraints in \eqref{multiple_prediction_regions} are valid, it suffices that at least one of them is satisfied by $x_{\tau|t}$. Hence, the dynamic agent constraints that are enforced in MPC are given by:
\begin{equation}\label{relaxed_CP_agent_constraints}
    \max_{0\leq s\leq t}\left\{c(x_{\tau|t},\Yhat_{\tau|s})- LC_{\tau|s}\right\}\geq0,
\end{equation}
for all $t$ and $\tau$. Figure~\ref{fig_CP_shrinking} shows that combining the constraints from \eqref{multiple_prediction_regions} into \eqref{relaxed_CP_agent_constraints} provides us with a valid constraint set for $x_{\tau|t}$, which expands the area of free space for the system at time $t$. Depending on the specific form of $c(\cdot,\cdot)$, an expression of \eqref{relaxed_CP_agent_constraints} without the maximum might be possible to derive (consider, e.g., the example in Section~\ref{Case_Study}). For complex constraints, simple underapproximations of the corresponding constraint sets could be considered.

We are now ready to present our model predictive controller. At each time $t$, our MPC scheme solves the following optimization problem: 
\begin{subequations}\label{MPC_problem}
\begin{align}
    \min_{\substack{x_{t|t},\ldots,x_{T|t} \\ u_{t|t},\ldots,u_{T-1|t}}}\; &J_f(x_{T|t})+\sum_{\tau=t}^{T-1}J(x_{\tau|t},u_{\tau|t}) \\
    \subto &x_{\tau+1|t} = f(x_{\tau|t},u_{\tau|t}), &\tau=t,\ldots,T-1 \\
        & x_{\tau+1|t}\in\calX_{\tau+1}, u_{\tau|t}\in \calU_{\tau}, &\tau=t,\ldots,T-1  \\
        &\max_{0\leq s\leq t}\{c(x_{\tau|t},\Yhat_{\tau|s})- LC_{\tau|s}\}\geq0,& \tau=t+1,\ldots,T\label{MPC_agent_constraints} \\
        &x_{t|t}=x_t,
\end{align}
\end{subequations}
where $u_{\tau|t}$ is the prediction of $u_{\tau}$ at time $t$. Once we obtain the optimal sequence $\{u_{t|t}^*,\ldots,$ $u_{T-1|t}^*\}$ of \eqref{MPC_problem}, the control input $u_t(x_t):=u_{t|t}^*$ is applied to system \eqref{system}. We repeat the same process at each time step, thus yielding a shrinking-horizon strategy. The full approach is described in Algorithm~\ref{algorithm}.

\begin{algorithm}[tbh]
\SetAlgoNoEnd
    \caption{Recursively Feasible Shrinking-Horizon MPC in Dynamic Environments with Conformal Prediction Guarantees}
    \label{algorithm}
    \KwInput{Mission time $T$, total number of data $K$, training set $D_{\train}:=\{Y_{0:T}^{(i)}:i\in\calI_{\train}\}$,\\ calibration set $D_{\calib}:=\{Y_{0:T}^{(i)}:i\in\calI_{\calib}\}$, trajectory predictor $\PREDICT$, failure probability $\delta$}
    \KwOutput{Closed-loop control inputs $u_0(x_0),\ldots,u_{T-1}(x_{T-1})$}
    \tcp{Offline computation of the conformal prediction values $C_{\tau|t}$}
    \For{$i=1,\ldots,K$}{
        \For{$t=0,\ldots,T-1$}{
            $(\Yhat_{t+1|t}^{(i)},\ldots,\Yhat_{T|t}^{(i)})\gets \PREDICT(Y_t^{(i)})$
        }
    }
    
    \For{$t=0,\ldots,T-1$}{
        \For{$\tau=t+1,\ldots,T$}{
        $\sigma_{\tau|t}\gets \max_{j\in\calI_{\train}}\norm{Y_{\tau}^{(j)}-\Yhat_{\tau|t}^{(j)}}$
        }
    }
    \For{$i\in\calI_{\calib}$}{
        $R^{(i)}\gets \max_{\substack{t=0,...,T-1\\\tau=t+1,...,T}}\{\norm{Y_{\tau}^{(i)}-\Yhat_{\tau|t}^{(i)}}/\sigma_{\tau|t}\}$
    }
    $R\gets\ceil{(|D_{\calib}|+1)(1-\delta)}$-th smallest value of the set $\{R^{(i)}: i\in\calI_{\calib}\}\cup\{\infty\}$
    
    \For{$t=0,\ldots,T-1$}{
        \For{$\tau=t+1,\ldots,T$}{
            $C_{\tau|t}\gets R \sigma_{\tau|t}$
        }
    }
    \tcp{Real-time MPC}
    \For{$t=0,\ldots,T-1$}{ 
        Observe $x_t$ and $Y_t$
        
        $(\Yhat_{t+1|t},\ldots,\Yhat_{T|t})\gets \PREDICT(Y_t)$
        
        Obtain the optimal control inputs $u_{t|t}^*,\ldots,u_{T-1|t}^*$ of \eqref{MPC_problem}
        
        $u_t(x_t)\gets u_{t|t}^*$
        
        Apply $u_t(x_t)$ to system \eqref{system}
    }
\end{algorithm}

\newpage
In the following theorem, we formalize the safety and recursive feasibility guarantees of the shrinking-horizon MPC scheme presented in Algorithm~\ref{algorithm}.
\begin{mytheorem}\label{theorem1}
Fix a failure probability\, $\delta\in(0,1)$. Let\, $\Yhat_{\tau|t}$\, and\, $C_{\tau|t}$\, be\, as\, in\, Lemma~\ref{lemma1}. Suppose the optimization problem \eqref{MPC_problem} is feasible at time step $t=0$. Then, applying the MPC scheme described in Algorithm~\ref{algorithm}, \eqref{MPC_problem} is feasible at every time step $t=0,\ldots,T-1$. Moreover, the resulting closed-loop trajectory ensures the satisfaction of the probabilistic safety guarantee \eqref{prob_agent_constraints}.
\end{mytheorem}

\begin{myremark}\label{remark3}
By construction, the dynamic agent constraints that we employ in our MPC scheme guarantee: i) recursive feasibility, and ii) probabilistic safety (see \eqref{prob_agent_constraints}). We emphasize that in our setting of shrinking-horizon MPC, recursive feasibility is not jeopardized by shortsightedness in the control design as in fixed-horizon MPC. However, it can be compromised when only the most recent prediction regions are incorporated in the design of safety constraints for MPC at each time step (consider, e.g., the shrinking-horizon MPC schemes presented in \citep{Lindemann2023a,Yu2023}). We ensure recursive feasibility via a gradual relaxation of the constraint \eqref{relaxed_CP_agent_constraints} with time $t$, which prevents solutions that are initially feasible from becoming infeasible later on (see the proof of Theorem~\ref{theorem1} in Appendix~\ref{proof_theorem1} for details). 
In Section~\ref{Case_Study}, we validate the above observations through a comparative case study. Probabilistic safety in our scheme is guaranteed by the fact that all constraints in \eqref{multiple_prediction_regions}, which are combined into the MPC constraint \eqref{relaxed_CP_agent_constraints}, correspond to valid prediction regions. This is essential given that the constraint \eqref{relaxed_CP_agent_constraints} holds if at least one of the constraints in \eqref{multiple_prediction_regions} is satisfied by $x_{\tau|t}$ (see Figure~\ref{fig_CP_shrinking}). By contrast, the conformal prediction regions from \citep{Lindemann2023a,Yu2023} are valid only for one-step ahead predictions (see Remark~\ref{remark1}), which implies that they would not ensure the guarantee \eqref{prob_agent_constraints} if incorporated in  \eqref{relaxed_CP_agent_constraints}. 
\end{myremark}

\section{Case Study: Navigation of a Mobile Robot around Pedestrians}\label{Case_Study}
In this section, we illustrate the efficacy of our model predictive controller in navigating a robot around pedestrians. Specifically, we consider a bicycle model \citep{Pepy2006}:
\begin{equation*}
\begin{bmatrix}p_{x,t+1}\\p_{y,t+1}\\\theta_{t+1}\\v_{t+1}  \end{bmatrix} = 
    \begin{bmatrix}
        p_{x,t}+\Delta v_t\cos{\theta_t}\\
        p_{y,t}+\Delta v_t\sin{\theta_t}\\
        \theta_{t}+\Delta\frac{v_t}{\ell}\tan{\phi_t}\\
        v_t+\Delta a_t
    \end{bmatrix},
\end{equation*}
where $p_t:=(p_{x,t},p_{y,t})$ is the position of the rear axle, $\theta_t$ is the orientation, $v_t$ is the velocity, $\ell:=0.5$ is the length, and $\Delta:=1/8$ is the sampling time. The control inputs are the steering angle $\phi_t\in\begin{bmatrix}-\pi/6,\pi/6\end{bmatrix}$ and the acceleration $a_t\in\begin{bmatrix}-5,5\end{bmatrix}$. We assume that the robot operates in an environment with three pedestrians. Let $Y_t\in\setR^6$ be the joint position of the pedestrians and $\varepsilon_{\textup{ped}}:=0.1$ the safety distance from each pedestrian. We are given a dataset $D$ of $2610$ pedestrian trajectories. Our goal is to navigate the robot from the initial point $p_0:=(3.5,-3)$ to the target point $p_{\textup{target}}:=(-1.8,1)$, while avoiding the three pedestrians. To achieve this, we employ the constraint $\norm{p_{20}-p_{\textup{target}}}_{\infty}\leq0.05$ and the cost function $J_{20}(x_0;u_0,\ldots,u_{19}):=\sum_{t=0}^{20}\norm{p_t-p_{\textup{target}}}_2^2$.\footnote{\label{footnote_exp}In this example, we have $x_t=\begin{bmatrix}p_{x,t},p_{y,t},\theta_t,v_t\end{bmatrix}^{\intercal}$ and $u_t=\begin{bmatrix}\phi_t,a_t\end{bmatrix}^{\intercal}$.} We also consider the objective \eqref{prob_agent_constraints} with $\delta=0.1$ and:
\begin{equation*}\label{constraint_function}
    c(x_t,Y_t) = \min_{j=1,2,3}\norm{p_t-Y_{t,j}}_{\infty}-\varepsilon_{\textup{ped}}-\ell.\footref{footnote_exp}
\end{equation*}
Note that the function $c(\cdot,\cdot)$ is $1$-Lipschitz in its second argument with respect to the norm $\norm{\cdot}_{\infty}$.

To demonstrate the benefits of our MPC algorithm, the MPC scheme from \citep{Lindemann2023a} is employed as benchmark. Implementation details can be found in Appendix~\ref{appendix_B}. Out of a total of $1000$ test trajectories, we found that $973$ are within our conformal prediction regions and $984$ are within the conformal prediction regions from \citep{Lindemann2023a}. We deduce that our approach results in tighter prediction regions, while maintaining the prediction guarantee \eqref{conformal_prediction_guarantee}. 

\begin{figure}[tbh]
   \centering
   \includegraphics[width=0.9\linewidth]{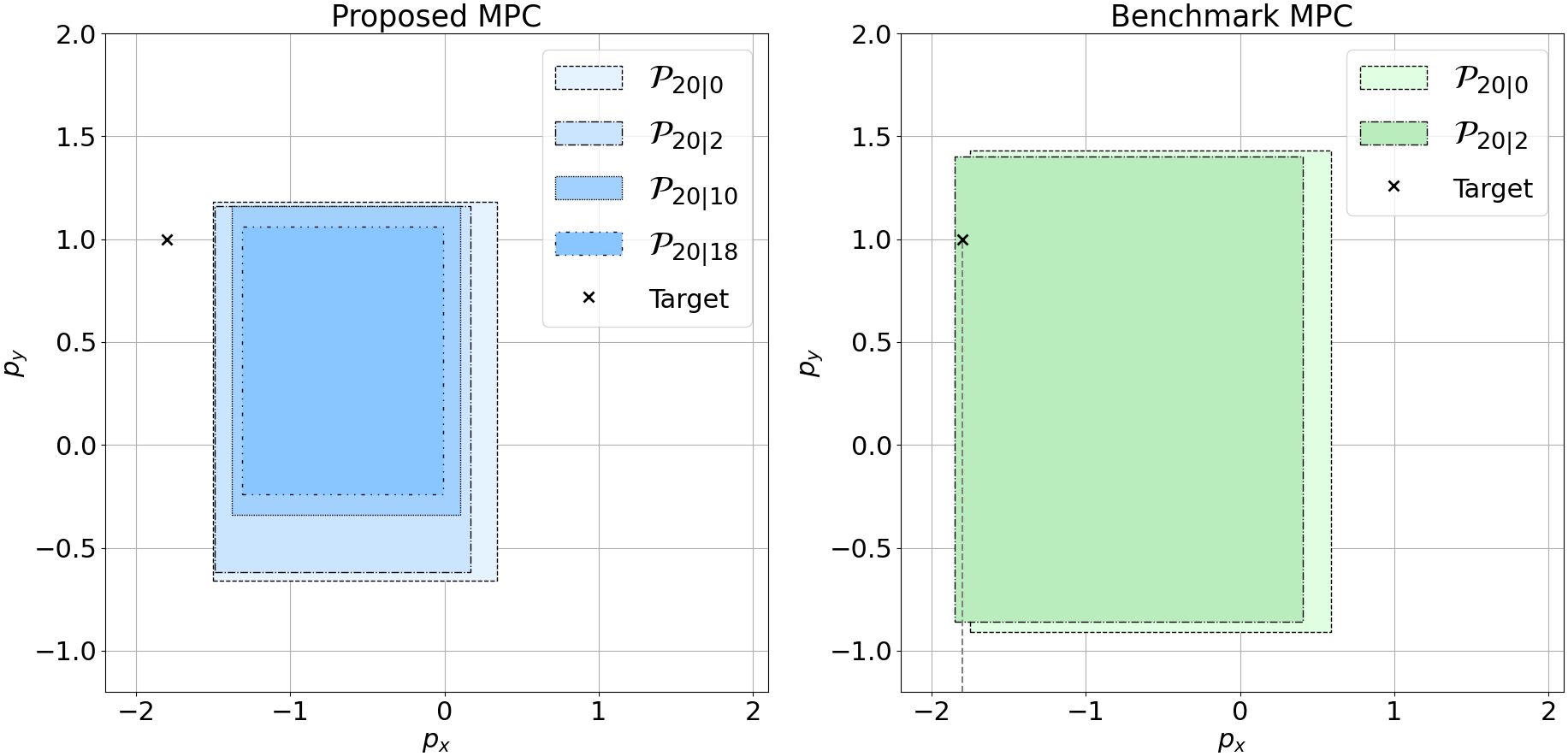}
   \caption{{\small Predicted unsafe areas corresponding to the final position of pedestrian $1$ at various times. In the proposed MPC, the target is deemed safe at all times. In the benchmark MPC, the target is deemed safe at time $0$ and unsafe at time $2$, rendering the controller recursively infeasible.}}
   
   \label{fig_shrinking_comparison}
\end{figure}

\begin{figure}[tbh]
\centering
\begin{minipage}{.47\linewidth}
  \centering
  \includegraphics[width=\linewidth]{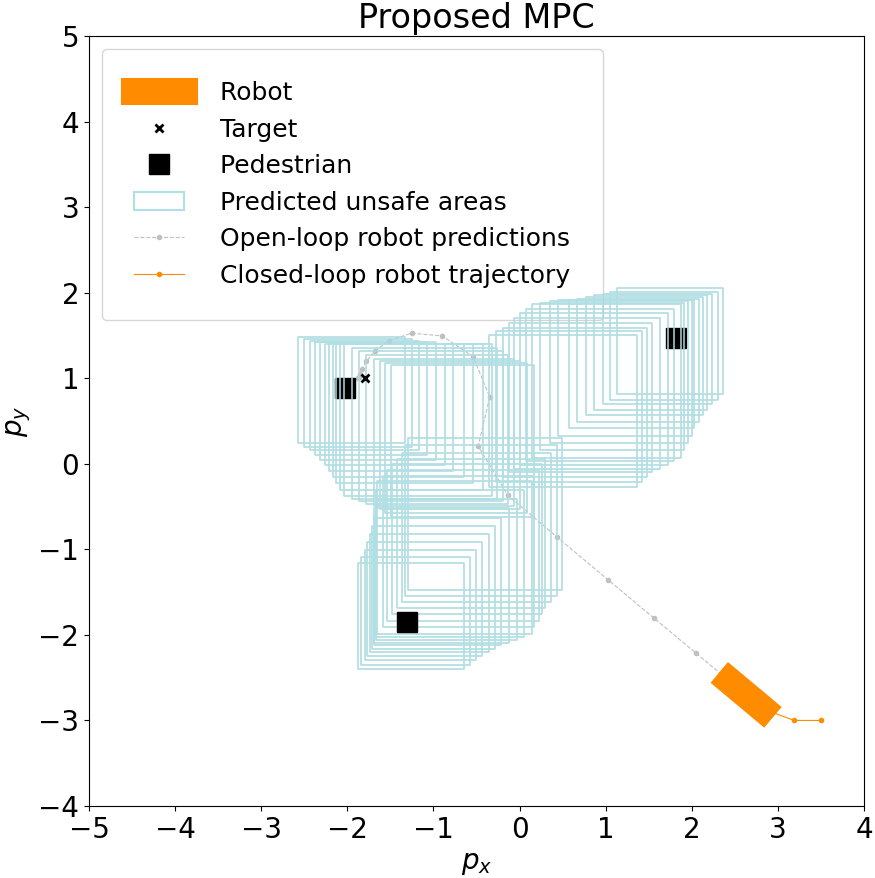}
  \label{fig:test1}
\end{minipage}
\hspace{-0cm}
\begin{minipage}{.47\linewidth}
  \centering
  \includegraphics[width=\linewidth]{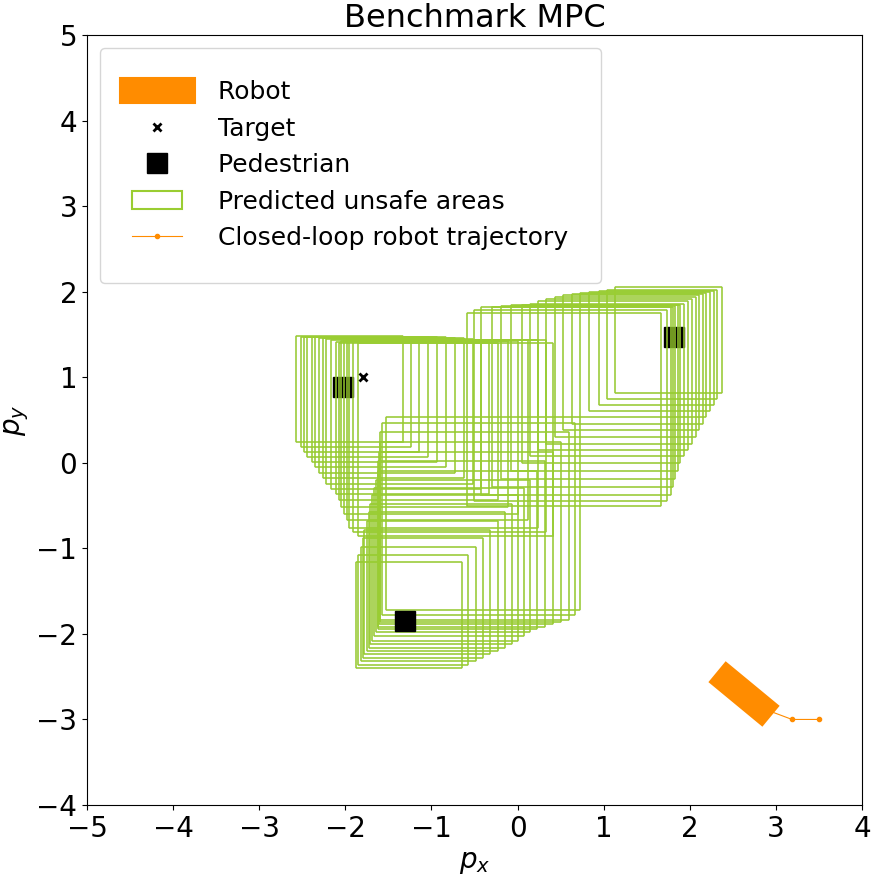}
  \label{fig:test2}
\end{minipage}
\caption{{\small Proposed and benchmark MPC at time $2$. For the proposed MPC, there exists a safe trajectory (gray line) that allows the robot to reach the target at time $20$. In contrast, the benchmark MPC becomes infeasible, as the target is predicted to be unsafe with respect to pedestrian $1$.}}
\label{mpc_comparison}
\end{figure}

We focus on a comparative simulation of the two MPC schemes, corresponding to one of the test trajectories. Let $\calP_{t}$ denote the unsafe area corresponding to pedestrian $1$ at time $t$ and let $\calP_{\tau|t}$ be the prediction of $\calP_{\tau}$ at time $t$, employed in the design of each MPC (see Appendix~\ref{appendix_B} for formal definitions). In Figure~\ref{fig_shrinking_comparison}, we draw: i) the unsafe areas $\calP_{20|t}$ of the proposed MPC at time steps $t=0,2,10,18$, and ii) the unsafe areas $\calP_{20|t}$ of the benchmark MPC at time steps $t=0,2$. We observe that the unsafe areas $\calP_{20|0}$ and $\calP_{20|2}$ are tighter in the proposed MPC than the benchmark MPC, which suggests a reduction of conservatism in our approach. In the proposed MPC, the target is consistently deemed safe at times $0$, $2$, $10$, and $18$, and recursive feasibility is verified in simulation. In the benchmark MPC, the target is deemed safe at time $0$ and unsafe at time $2$, rendering the controller recursively infeasible. This inconsistency results from the fact that the benchmark scheme overlooks $\calP_{20|0}$ in the design of $\calP_{20|2}$. In contrast, the constraint \eqref{relaxed_CP_agent_constraints} of the proposed scheme guarantees that the unsafe areas $\calP_{20|t}$ consistently shrink, by accounting for all available predicted unsafe areas $\calP_{20|0},\ldots,\calP_{20|t-1}$ in their design (see Remark~\ref{remark3}). Animations for both schemes can be found at \href{https://tinyurl.com/yckdzvz2}{\color{blue}{https://tinyurl.com/yckdzvz2}}. Figure~\ref{mpc_comparison} shows the frames at time $2$. 

\section{Future Work}
Moving forward, our paper opens up several research directions. First, 
in order to improve the statistical efficiency of our conformal prediction method, we could leverage results for dependent data to analyze each trajectory. Moreover, it would be interesting to explore how changes in the distribution of the dynamic agents can be incorporated in our approach. Finally, a scalable extension of our scheme to settings with an arbitrary number of agents could be considered.

\section*{Acknowledgements}
Charis Stamouli and George J. Pappas acknowledge support from NSF award SLES-2331880.

\bibliography{arxiv_version}

\appendix
\section{Proofs}\label{appendix_A}
\subsection{Proof of Lemma~\ref{lemma1}}
Since the calibration trajectories $Y^{(i)}\in D_{\calib}$ are i.i.d., the conformity scores \eqref{conformity_scores} are i.i.d., conditioned on the dataset $D_{\train}$. Given that exchangeability is a strictly weaker condition than independence and identical distribution \citep{Vovk2005}, from \citep[Lemma 1]{Tibshirani2019} we obtain the conditional property:
\begin{equation}\label{lemma1_1}
    \Prob\left(\max_{\substack{t=0,...,T-1\\\tau=t+1,...,T}}\left\{\frac{\norm[\big]{Y_{\tau}-\Yhat_{\tau|t}}}{\sigma_{\tau|t}}\right\}\leq R\,\Bigg|\,Y^{(i)},i\in\calI_{\train}\right)\geq1-\delta,
\end{equation}
where $R$ is defined as in the lemma statement. By marginalizing over the dataset $D_{\train}$, \eqref{lemma1_1} yields the unconditional property:
\begin{equation}\label{lemma1_2}
    \Prob\left(\max_{\substack{t=0,...,T-1\\\tau=t+1,...,T}}\left\{\frac{\norm[\big]{Y_{\tau}-\Yhat_{\tau|t}}}{\sigma_{\tau|t}}\right\}\leq R\right)\geq1-\delta.
\end{equation}
The guarantee \eqref{conformal_prediction_guarantee} directly follows from rewriting \eqref{lemma1_2} as:
\begin{equation*}
    \Prob\left(\bigcap_{t=0}^{T-1}\bigcap_{\tau=t+1}^T\left\{\norm[\big]{Y_{\tau}-\Yhat_{\tau|t}}\leq R \sigma_{\tau|t}\right\}\right)\geq1-\delta
\end{equation*}
and setting $C_{\tau|t}=R\sigma_{\tau|t}$, $\forall t,\tau$.

$\qed$

\subsection{Proof of Theorem~\ref{theorem1}}\label{proof_theorem1}
Assume that \eqref{MPC_problem} is feasible at time step $t=0$. We will prove that the solution of \eqref{MPC_problem} guarantees recursive feasibility, that is, we will show that \eqref{MPC_problem} is feasible at every time step $t=0,\ldots,T-1$. Let $\{u_{0|0}^*,\ldots,u_{T-1|0}^*\}$, $\{x_{0|0},\ldots,x_{T|0}\}$ denote the optimal control and state sequence, respectively, at time step $t=0$. We apply the control input $u_0(x_0):=u_{0|0}^*$ to system \eqref{system} and thus the state evolves to $x_{1|0}$. We will show that at time step $t=1$ the control sequence $\{u_{1|0}^*,\ldots,u_{T-1|0}^*\}$ and the state sequence $\{x_{1|0},\ldots,x_{T|0}\}$ compose a feasible solution of \eqref{MPC_problem}. Let $\calS_{\tau|t}$ denote the set defined by the constraint \eqref{relaxed_CP_agent_constraints}. From \eqref{relaxed_CP_agent_constraints} we deduce that $\calS_{\tau|0}\subseteq\calS_{\tau|1}$, for all $\tau=2,\ldots,T$. Hence, since $x_{2|0}\in\calS_{2|0},\ldots,x_{T|0}\in\calS_{T|0}$, we conclude that $x_{2|0}\in\calS_{2|1},\ldots,x_{T|0}\in\calS_{T|1}$. It is also clear that $x_{2|0}\in\calX_2,\ldots,x_{T|0}\in\calX_T$ and $u_{1|0}^*\in\calU_1,\ldots,u_{T-1|0}^*\in\calU_{T-1}$. Therefore, we deduce that \eqref{MPC_problem} is feasible at time step $t=1$. Similarly, we can prove that \eqref{MPC_problem} is feasible at time steps $t=2,\ldots,T-1$, thus completing the proof of recursive feasibility. 

By construction, the constraints \eqref{MPC_agent_constraints}, which are employed in the MPC scheme of Algorithm~\ref{algorithm}, ensure the satisfaction of \eqref{prob_agent_constraints} (see the derivation of \eqref{relaxed_CP_agent_constraints} in Section~\ref{MPC_in_Dynamic_Environments_via_Conformal_Prediction}).

$\qed$

\section{Experiment Details}\label{appendix_B}
In this section, we present the implementation details for the two MPC schemes discussed in Section~\ref{Case_Study}. Moreover, we provide formal definitions for the sets $\calP_{\tau|t}$ depicted in Figure~\ref{fig_shrinking_comparison}.

For the data generation, we used the ORCA simulator \citep{Van2011} in TrajNet++ \citep{Kothari2021}. We split the data into: i) a training set of size $2000$, ii) a calibration set of size $610$, and iii) a test set of size $1000$. For trajectory prediction, we use the social Long Short-Term Memory (LSTM) model from \citep{Alahi2016}. The model features one vanilla LSTM for each human. Shared weights are leveraged to model social interactions. For simplicity, initial observations $Y_{-19},\ldots,Y_0$ are assumed. We train a social LSTM with three individual LSTMs of depth $d:=128$. The learned model and the three datasets are employed in the implementation of Algorithm~\ref{algorithm} and \citep[Algorithm 1]{Lindemann2023a}.

Recall that the function $c(\cdot,\cdot)$ defined in Section~\ref{Case_Study} is $1$-Lipschitz with respect to the norm $\norm{\cdot}_{\infty}$. At each time $t$, the dynamic agent constraints of the benchmark MPC are given by:
\begin{equation}\label{benhcmark_constraints}
    c(x_{\tau|t},\Yhat_{\tau|t})\geq \widetilde{C}_{\tau|t},\;\tau=t+1,\ldots,T.
\end{equation}
The design of \eqref{benhcmark_constraints} relies on conformal prediction regions of the form:
\begin{equation*}\label{bench_prediction_regions}
        \norm[\big]{Y_{\tau}-\Yhat_{\tau|t}}_{\infty}\leq \widetilde{C}_{\tau|t},
\end{equation*}
where the values $\widetilde{C}_{\tau|t}$ are computed based on separate conformity scores $R_{\tau|t}^{(i)}$, for each pair $(t,\tau)$ (see \citep{Lindemann2023a} for details). For clarity, we formally define the sets $\calP_{\tau|t}$ depicted in Figure~\ref{fig_shrinking_comparison}. In the proposed MPC, we have:
\begin{equation}\label{P_set}
    \calP_{\tau|t}=\left\{p_{\tau|t}\in\setR^2: \max_{s=0,\ldots,t}\left\{\norm[\big]{p_{\tau|t}-\Yhat_{\tau|s,1}}_{\infty}-\varepsilon_{\textup{ped}}-\ell-C_{\tau|s}\right\}<0 \right\},
\end{equation}
while in the benchmark MPC, we have:
\begin{equation*}
    \calP_{\tau|t}=\left\{p_{\tau|t}\in\setR^2: \norm[\big]{p_{\tau|t}-\Yhat_{\tau|t,1}}_{\infty}-\varepsilon_{\textup{ped}}-\ell-\widetilde{C}_{\tau|t}<0 \right\}.
\end{equation*}
We note that a description of the constraint set \eqref{P_set} without the maximum is straightforward to derive.

\end{document}